\documentclass[conference]{IEEEtran}
\IEEEoverridecommandlockouts
\usepackage{cite}
\usepackage{amsmath,amssymb,amsfonts}
\usepackage{algorithmic}
\usepackage{makecell}
\usepackage[hidelinks]{hyperref}
\usepackage{graphicx}
\usepackage{textcomp}
\usepackage{tabularx}
\usepackage{balance}
\usepackage{xcolor}
\usepackage{tcolorbox}
\usepackage{url}
\def\BibTeX{{\rm B\kern-.05em{\sc i\kern-.025em b}\kern-.08em
    T\kern-.1667em\lower.7ex\hbox{E}\kern-.125emX}}
\begin{document}

\title{DocFetch - Towards Generating Software Documentation from Multiple Software Artifacts}

\author{\IEEEauthorblockN{Akhila Sri Manasa Venigalla and Sridhar Chimalakonda}
\IEEEauthorblockA{\textit{Research in Intelligent Software \& Human Analytics (RISHA) Lab}\\
\textit{Department of Computer Science and Engineering} \\
\textit{Indian Institute of Technology Tirupati}\\
Tirupati, India\\
\{cs19d504, ch\}@iittp.ac.in }
}
\maketitle

\begin{abstract}
  Software Documentation plays a major role in the usage and development of a project. Widespread adoption of open source software projects contributes to larger and faster development of the projects, making it difficult to maintain the associated documentation. 
Existing automated approaches to generate documentation largely focus on source code. However, information useful for documentation is observed to be scattered across various artifacts that co-evolve with the source code. Leveraging this information across multiple artifacts can reduce the effort involved in maintaining documentation. 
Hence, we propose \textit{DocFetch}, to generate different types of documentation from multiple software artifacts. We employ a multi-layer prompt based LLM and generate structured documentation corresponding to different documentation types for the data consolidated in \textit{DocMine} dataset. 
We evaluate the performance of \textit{DocFetch} using a manually curated groundtruth dataset by analysing the artifacts in \textit{DocMine}. The evaluation yields a highest BLEU-4 score of 43.24\%  and ROUGE-L score of 0.39 for generation of \textit{api-related} and \textit{file-related} information from five documentation sources. The generation of other documentation type related information also reported BLEU-4 scores close to 30\% indicating good performance of the approach. 
Thus, \textit{DocFetch} can be employed to semi-automatically generate documentation, and helps in comprehending the projects with minimal effort in maintaining the documentation. The tool and the results of experimentation are available \underline{\href{https://drive.google.com/drive/folders/1sEyjB9TFRZNlC7FY-11khpwk3c5gTf9P?usp=sharing}  {here}}.
\end{abstract}

\begin{IEEEkeywords}
Software Documentation, LLMs, Commits, Issues, Pull Requests
\end{IEEEkeywords}

\section{Introduction}
\label{intro}
Growing adoption of open source software (OSS) platforms facilitates contributions from wider developer community across the world. For instance, popular projects on GitHub, such as \textit{microsoft/vscode}\footnote{\url{https://github.com/microsoft/vscode}} (173K stars) and \textit{tensorflow/tensorflow}\footnote{\url{https://github.com/tensorflow/tensorflow}} (190K stars) receive contributions from 4.3K and 7.3K contributors respectively as of May 2025. Availability of open source projects further contributes to increased reuse of the projects, thereby reducing the development effort. The \textit{tensorflow/tensorflow} repository alone has about 216 releases and is used by more than 519K developers as of May 2025, indicating wide reuse.
Software Documentation supports software practitioners with varied roles, in development and maintenance of the projects \cite{garousi2015usage, alamin2023developer}. 
Good quality documentation facilitates better comprehension of software projects \cite{borges2018s}, thereby supporting optimal reuse source projects \cite{gao2025adapting,liu2021reproducibility, mitchell2022} and attracting contributions to open source projects \cite{ehsan2020empirical,xiao2023early, bao2019large}.

Analysis of about 290K GitHub projects and survey with 148 developers highlighted that detailed and well-maintained documentation is an important factor in promoting sustained long-term activity and contributor retention \cite{xiao2023early, bao2019large}.
For instance, GitHub repositories maintaining consistent readme files (widely considered as an important documentation) with frequent changes received higher popularity than others \cite{venigalla2025there, aggarwal2014co}. Further, better organised readme files also attracted higher popularity to the repositories \cite{venigalla2025there, wang2023study, fan2021makes}.
 
But, the documentation in its current state often suffers from issues such as incompleteness, inconsistencies, insufficiency, incorrectness and so on \cite{aghajani2019software, wu2024comprehensive, rong2024code}.
While constant updation could solve majority of the challenges associated with documentation, it is difficult to
manually document the updates of faster and ever-evolving projects \cite{huang2024empirical,liu2020atom}.
Collaborative software development facilitated by OSS platforms supports faster progress of the projects. While faster development is desirable, the increased rate of development also influences the quality of documentation and poses challenges to ensure well-updated documentation. For instance, \textit{tensorflow/tensorflow} repository on GitHub receives an average of 300 commits per week. Documenting every change caused due to these commits is time-taking and effort-intensive, thus making it challenging to document the updates. 

To facilitate faster documentation in such contexts, researchers have proposed several automated approaches to generate documentations such as source code summaries and commit messages. 
A number of neural networks models including \textit{seq2seq} architectures and hierarchical attention networks, alongside other transformer models, 
were proposed for the automatic generation of code summaries, considering both linear and structural representations of the source code as input \cite{wei2020retrieve, bansal2023function, gao2023code}. Similar models considering different aspects of the source code such as call graphs were used to generate commit messages based on the changes made to the code \cite{liu2020atom, eliseeva2023commit, huang2024empirical}. Researchers have also explored the use of pre-trained large language models that accept source code as input, to summarize code and its changes, for better understanding of the source code \cite{li2024understanding, khan2022automatic, nam2024using}. 

Other artifacts apart from source code, associated with the projects also co-evolve with source code. These artifacts such as \textit{commits}, \textit{pull-requests} and \textit{issues} were explored in the literature to obtain insights into various aspects of the projects such as contributor characteristics \cite{jamieson2024predicting}, fault-detection \cite{kuramoto2022visual}, developer on-boarding \cite{xiao2022recommending}, security vulnerabilities \cite{mazuera2022taxonomy} and so on.
Integrating and analysing multiple types of artifacts provides better insights for certain tasks. 
Different combinations of pull-requests, issues and commits were analysed to identify bots in GitHub \cite{golzadeh2021ground}, locate features in source code \cite{kruger2019my}, visualize history and nature of issues \cite{fiechter2021visualizing}, and to generate pull request titles \cite{zhang2022automatic}. 
These artifacts also contain information corresponding to different types of documentation \cite{venigalla2024exploratory}, indicating the scattered nature of documentation across different project artifacts. 

In this work, we aim to identify and integrate documentation-related information to generate different types of documentation.
We develop \textit{DocFetch}, a research prototype to demonstrate the generation of  desired documentation type, using an approach that employs pre-trained Large Language Models (LLMs).
We consider the documentation sources and types discussed in \cite{venigalla2024exploratory} and utilize the information extracted from these sources, consolidated in the \textit{DocMine} \cite{venigalla2023docmine} dataset to generate documentation for the projects.

Pre-trained LLMs are being widely used to generate various aspects of software projects, including source code \cite{du2024evaluating, liu2024no, mathews2024test}, code changes \cite{li2024understanding, wu2025empirical}, source code summaries \cite{ahmed2022few, ahmed2024automatic}, and test cases \cite{wang2024hits, ouedraogo2024llms}.
The summaries of source code generated using LLMs increased the quality of software documentation \cite{hou2024large}. 
We leverage the advantage of employing few-shot prompts to guide pre-trained LLMs in generating documentation that adheres to a predetermined structural format. We employ pre-trained LLMs at multiple levels to process each of the documentation sources and to subsequently integrate the generated information across the sources.

We evaluated the information generated by \textit{DocFetch} from each documentation source, for different documentation types against a manually curated ground-truth dataset. The results of the evaluation indicate high degree of similarity with the groundtruth dataset, with an average BLEU score of 30.2 and ROUGE score of 0.38. These scores are 
comparable to majority of the metric scores achieved in the state-of-the-art code summarization tasks \cite{ahmed2024automatic} that employ pre-trained LLMs for generation of code comments and summaries.


\section{Background}
\label{background}
In this section, we discuss the relevant context associated with different documentation sources and types, pre-trained LLMs and few-shot prompts.

\subsection{Documentation Types and Sources}

Software artifacts associated with the projects co-evolve with source code and contain rich textual 
information which can be leveraged to comprehend about the project. The content present in these artifacts can contain information useful to enhance or generate documentation. \cite{venigalla2024exploratory} have conducted a large scale exploratory study of 1.38 million artifacts across 950 GitHub repositories to identify such artifacts which can be considered as documentation sources. They also classified the documentation into six different types, based on qualitative analysis methods. We briefly discuss the documentation types and sources proposed by \cite{venigalla2024exploratory} below.

\textit{\textbf{Documentation Types - }} \textit{A documentation type is determined based on the content it contains, providing insights into various aspects of a project. } 
\cite{venigalla2024exploratory} propose the following six documentation types:
\begin{enumerate}
    \item \textit{File-related Documentation Type.} This type of documentation discusses information corresponding to the dependencies, organisation and updates associated with various files in the project.

    \item \textit{Error/Bug-related Documentation Type.} It presents instances of various errors that occur across the projects and the corresponding measures taken to resolve these errors.

    \item \textit{Project-related Documentation Type.} This type of documentation corresponds to the project-level information such as the updates made at project level, environments supported by the project, instructions associated with using and contributing to the project.

    \item  \textit{API-related Documentation Type.} It comprises the information associated with different APIs used and defined in the project, associated arguments and return types.

    \item \textit{License-related Documentation Type.} It discusses the licenses under which the project is made available and other permissions provided by the project.

    \item \textit{Architecture-related Documentation Type.} This type of documentation corresponds to the organisation of the project, different modules into which the project is organized and the interactions among these modules, usually represented pictorially.
\end{enumerate}

\textit{\textbf{Documentation Sources -}}  \textit{A documentation source is a software artifact within a project that holds the potential to offer information contributing to an improved understanding of the project.} In their exploratory study, \cite{venigalla2024exploratory} explore the distribution of different documentation types across the following documentation sources. 
\begin{enumerate}
    \item \textit{Commits}.  15.6\% of the extracted information from commits corresponded to file-related, 31.5\% to error-related, 39\% to project related, 2.7\% to API-related and 3.9\% to License-related information.

    \item \textit{Issues}. Among the total information extracted from Issues, 17.2\% contained file-related documentation, 35.5\% error-related, 24.2\% project related, 4\% API-related and 3.4\% License-related information.

    \item \textit{Pull Requests}. The text extracted from Pull requests comprised 16.34\% of file-related information, 36.7\% of error-related information, 26.9\% of project related, 4.58\% of API-related and 4.43\% of License-related information. 

    \item \textit{Textual Files}.
14.8\% of the extracted information from textual files corresponded to file-related, 16.32\% to error-related, 29.74\% to project related, 2.94\% to API-related and 15.28\% to License-related information.

    \item \textit{Source code Comments}.  Among the total content extracted from source code comments, 11.82\% contained file-related documentation, 7.66\% error-related,20.54\% project related, 11.6\% API-related and 29.16\% License-related information.    
\end{enumerate}
The text extracted from 1.38M artifacts across 950 GitHub repositories, from documentation perspective is consolidated in the \textit{DocMine} dataset \cite{venigalla2023docmine}, which is leveraged to design and develop \textit{DocFetch}.

\subsection{Pre-trained Large Language Models}
Large Language Models (LLMs) are designed based on the transformer architecture, with the transformer's self attention module as the basic building block of the LLMs \cite{chang2024survey}. They contain billions of parameters and are pre-trained on large textual corpora to support various Natural Language Processing (NLP) tasks such as generation, summarisation and translation of text \cite{du2024evaluating}. Pre-trained LLMs reduce the requirement for expensive resources associated with high computational power and memory, while also overcoming the need to create labelled datasets for specific tasks \cite{ahmed2024automatic}.  Pre-trained LLMs such as BERT, GPT, RoBERTa and Gemini are being used for various NLP tasks, including those specific to computer programming.
They are widely being used for various downstream software engineering tasks including fixing bugs \cite{xia2023automated}, code generation and summarisation \cite{nam2024using, du2024evaluating, ahmed2024automatic, macneil2022generating} and testcase generation \cite{gu2023llm, dakhel2024effective}. 

Fine-tuning these LLMs by adding domain-specific information further enhances the performance of these models. The in-context learning feature of LLMs facilitates the model to be trained with minimal data points to generate text based on a given context or prompt \cite{ahmed2024automatic}. These prompts support more context towards obtaining precise results corresponding to the specific needs, thus reducing the efforts vested in further processing of the obtained information \cite{nam2024using}. Prompting an LLM involves a text description of the task to be addressed, which is passed to the model as input. The model generates desired text based on the description passed as input. 

These textual descriptions passed as prompts can also involve examples demonstrating the sample output for a sample input \cite{ahmed2024automatic}. This process of passing examples as prompts helps in providing context to the LLMs and also facilitates fine-tuning of the LLM. 
Each input-output example pair refers to a shot, and the number of such examples determine the number of shots used to learn the context. Passing zero examples refers to \textit{zero-shot} learning, while passing one example refers to \textit{one-shot} learning. Passing more than one example refers to \textit{few-shot} learning \cite{khan2022automatic}. This process of enabling one-shot and few-shot learning have proved to give fine-tuned results in desired formats \cite{ahmed2024automatic}. 
Hence, in our approach of generating documentation, we employ the pre-trained LLM available for use through API with few-shot learning. 

\section{Methodology}
\label{meth}
We design \textit{DocFetch} in three main phases - (i) Groundtruth Creation, (ii) Prompt Design and (iii) Architecture Design. 
\textit{DocFetch} is presented as a react-based application that takes inputs from the user corresponding to the (a) repository of interest and (b) desired documentation type for which the user expects information. \textit{DocFetch} then provides the  user with a downloadable `.json' file corresponding to the desired documentation type.
We discuss the details of each of these phases in this section.

\subsection{Groundtruth Creation}
We consider the artifacts consolidated in the \textit{DocMine} dataset to create the groundtruth and to build \textit{DocFetch} tool. Majority of the approaches in the literature focus on generating summaries from source code or explore source code and associated  messages to generate textual content such as commit messages and pull request titles and descriptions. As a result, the datasets available in the literature to analyse performance of documentation generation tools largely comprise source code and the associated comments or the commit messages or corresponding pull request descriptions. 

While different documentation types such as file-related, error/ bug-related, API-related and so on are proposed, there does not exist any dataset that comprises of manually written content corresponding to these documentation types. \textit{DocMine} dataset contains the distribution of these documentation types among different software artifacts, but does not specifically extract and validate the content corresponding to these documentation types from the sources. Due to this unavailability 
of the datasets to evaluate \textit{DocFetch}, we manually analyse the artifacts in \textit{DocMine} dataset and create a groundtruth dataset comprising of the artifact and the information corresponding to each documentation type that can be deduced from the artifact.

The creation of groundtruth dataset requires multiple iterations of manual analysis of the artifacts consolidated in the \textit{DocMine} dataset. Manually analysing all the 1.35M artifacts is time consuming and effort intensive. Hence, we select at least a significant sample of these artifacts, with 99\% confidence and 5\% margin of error, which accounts to 664 artifacts. 
We consider 1205 artifacts from 18 repositories, which is greater than the significant sample.

Each of these artifacts are carefully analysed by the first author of this paper to identify different aspects being discussed in them, corresponding to each documentation type. To ensure efficient comparison of the results with the groundtruth dataset, we design a structured format for each documentation type based on the documentation source from which information is analysed. This helps us in providing structure to the unstructured textual artifacts. 
We curate the groundtruth from each of the artifacts across the 18 repositories for each documentation type in alignment to this pre-defined structure, and store it in `.json' format. 

\begin{figure}[h!]
    \centering
    \includegraphics[width = \linewidth]{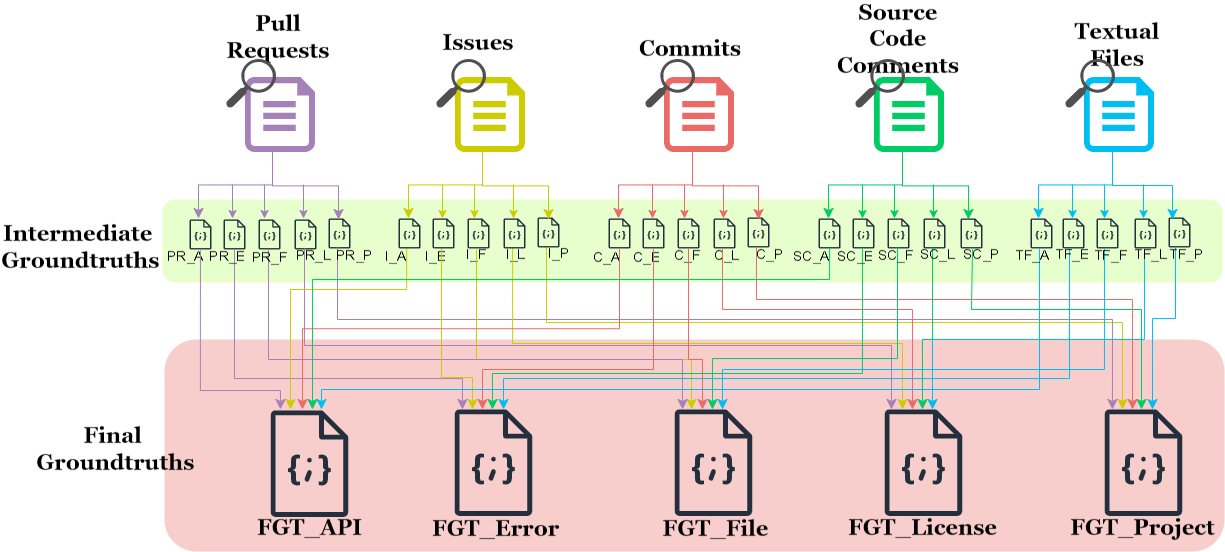}
    \caption{Groundtruth creation process, where X\_Y indicates \textit{intermediate groundtruth} of Y-related documentation obtained from documentation source X and FGT\_Y indicates \textit{final groundtruth} of Y-related documentation. For instance, PR\_A indicates \textit{intermediate groundtruth} of API-related documentation obtained from pull-requests and FGT\_API indicates \textit{final groundtruth} of API-related documentation consolidated from all documentation sources. }
    \label{fig:gt_creation}
\end{figure}

As a result, we create one groundtruth from each of the documentation sources for every documentation type, i.e., for a repository `X', we curated groundtruth from pull-requests, commits, issues, textual files and source code comments for api-related documentation. 
This accounts to 5 groundtruths from the documentation sources corresponding to every documentation type, thus accounting to 25 (5 documentation sources x 5 documentation types) groundtruths for every repository. We refer to these 25 groundtruths as \textit{intermediate groundtruths}. 

Further, we consolidate the information in each of the five
ground-truths corresponding to a documentation type, to arrive at the final groundtruth for the documentation type from all the documentation sources of the repository. Finally, each repository is associated with 30 (25 intermediate groundtruths + 5 final groundtruths) groundtruths. Fig. \ref{fig:gt_creation} depicts the groundtruth creation process, with intermediate groundtruths and final groundtruths obtained for each documentation type from all the documentation sources for any repository under consideration.

Manual curation of groundtruths consumed an average of 20 minutes each, accounting to a total of about 10,800 minutes (180 man hours). These groundtruths were then reviewed by two volunteer researchers, a PhD scholar and a Masters student to verify the correctness of the manually written documentation type information from each of the documentation sources. Any discrepancies in the groundtruth content was resolved through repeated discussions among the first author and the volunteer researchers. Finally, after two rounds of scrutiny, we achieved 100\% rate of agreement among the author and the volunteers.

\subsection{Prompt Design} 
We consider the pre-trained LLM model, \texttt{gemini-2.0-flash-lite-preview-02-05}, hereafter, referred to as \texttt{GFLP2.0}, in designing \textit{DocFetch} as it facilitates larger context window size of 37K characters. The documents we pass as inputs to the LLMs comprise information encompassing the project history and hence are usually large in size. Hence, a pre-trained LLM with larger context window size, as in case of \textit{GFLP2.0} is suitable for our context of generating documentation type information from the documentation sources.
We initially prompted the LLM model to list discussions related to a specific documentation type in a given input. 


For example, to generate \textit{error-related} documentation from pull-requests, the prompt to the LLM is passed as:

\begin{tcolorbox}[  colback=gray!5!white,colframe=gray!75!black,
  fonttitle=\bfseries, title= ]
`\textit{list out all discussions related to errors in $<$pull request text$>$ and explain the respective discussions}'. 
\end{tcolorbox}

While such prompts generated useful and appropriate, the generated outputs did not have any consistent structure, which made it challenging to evaluate the outputs on a large scale against any of the metrics to assess quality of the generated outputs. Hence, we decided to modify the prompts to generate a structured output. Further, passing one-shot and few-shot prompts to the model has proven to be beneficial while using pre-trained LLM models for generating code summaries \cite{khan2022automatic, ahmed2022few, ahmed2024automatic}. Hence, we employ two-shot and one-shot prompts to obtain outputs in desired format.

To ensure that the output of the pre-trained model follows the pre-defined structure for every documentation type from each documentation source, we employ two-shot learning. 
We pass two sample input and output pairs as a prompt to the model. With these input-output pairs, we attempt to teach the model the expected output structure. Hence, we prompt the model to learn the analogy in the examples and to generate the output for a new input. The input is the content extracted from the documentation source of the repository and the output is the desired documentation type following the predefined structure (json format).

\begin{figure}[h!]
    \centering
    \includegraphics[width = \linewidth]{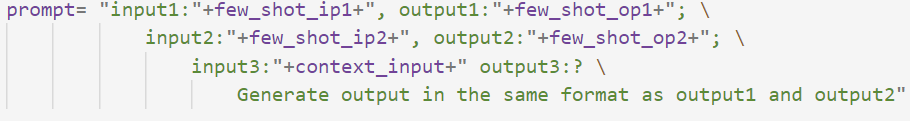}
    \caption{Example of the prompt used to the model, where `few\_shot\_ip1' and `few\_shot\_ip2' correspond to sample inputs, `few\_shot\_op1' and `few\_shot\_op2' correspond to the respective outputs of the sample inputs and   `context\_input' corresponds to the documentation source under consideration.}
    \label{fig:prompt}
\end{figure}

\begin{figure}[!h]
    \centering
\includegraphics[width = 0.8\linewidth]{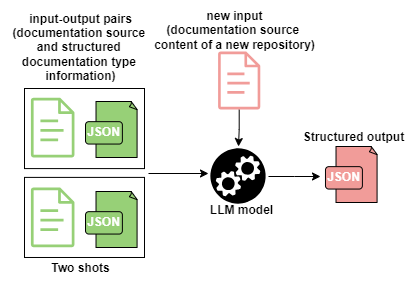}
    \caption{Level 1 architecture to generate documentation-type information in \textit{.json} format from a specific documentation source.}
    \label{fig:level1}
\end{figure}

We use the ground truths designed in the previous phase as prompts to the model, as illustrated in Fig.\ref{fig:prompt}.
For every pair of  documentation type and documentation source, the corresponding \textit{intermediate} groundtruths are passed as the sample input-output pairs. 
The output generated is stored as a `.json' file. The architecture of passing the prompt and obtaining the output is presented in Fig. \ref{fig:level1}.

Since the input-output pairs used as shots belong to the ground-truth dataset, which is manually curated, we attempt to teach the context to the model through these shots. The outputs in these pairs are carefully curated to include the details of the desired documentation type present in the corresponding inputs. For instance, the shots passed in the prompt to obtain `error-related' information from pull-requests of a specific repository include the pull request text extracted from two random repositories in the \textit{DocMine} dataset as inputs and the corresponding `error-related' information observed in the corresponding pull-request in structured format as respective outputs. Thus, the context of the documentation type is inherently provided to the model in the form of shots in the prompt, thus facilitating in-context learning.

\subsection{Architecture Design} 
We divide the design of \textit{DocFetch} tool broadly into two levels. We design the first level to generate documentation-type related information from each of the documentation sources individually. We then consolidate this information across all the documentation sources in the second level. We discuss the architecture of \textit{DocFetch} in detail through Fig. \ref{fig:arch} below. 

Based on the documentation type of interest and desired repository, we select the \textit{intermediate} groundtruths of  the documentation sources -  pull-requests, issues, commits, source code comments and textual files corresponding to the documentation type of interest, associated with any two repositories in the \textit{DocMine} dataset. 
In level 1, the data corresponding to each documentation source is individually passed to the LLM model to generate structured output for the desired documentation type. As a result, we employ five LLM models in parallel in level 1, one for each documentation source. We finally obtain five different outputs, one from each of the LLM, corresponding to every documentation source. 

\begin{figure}[h!]
    \centering
    \includegraphics[width = \linewidth]{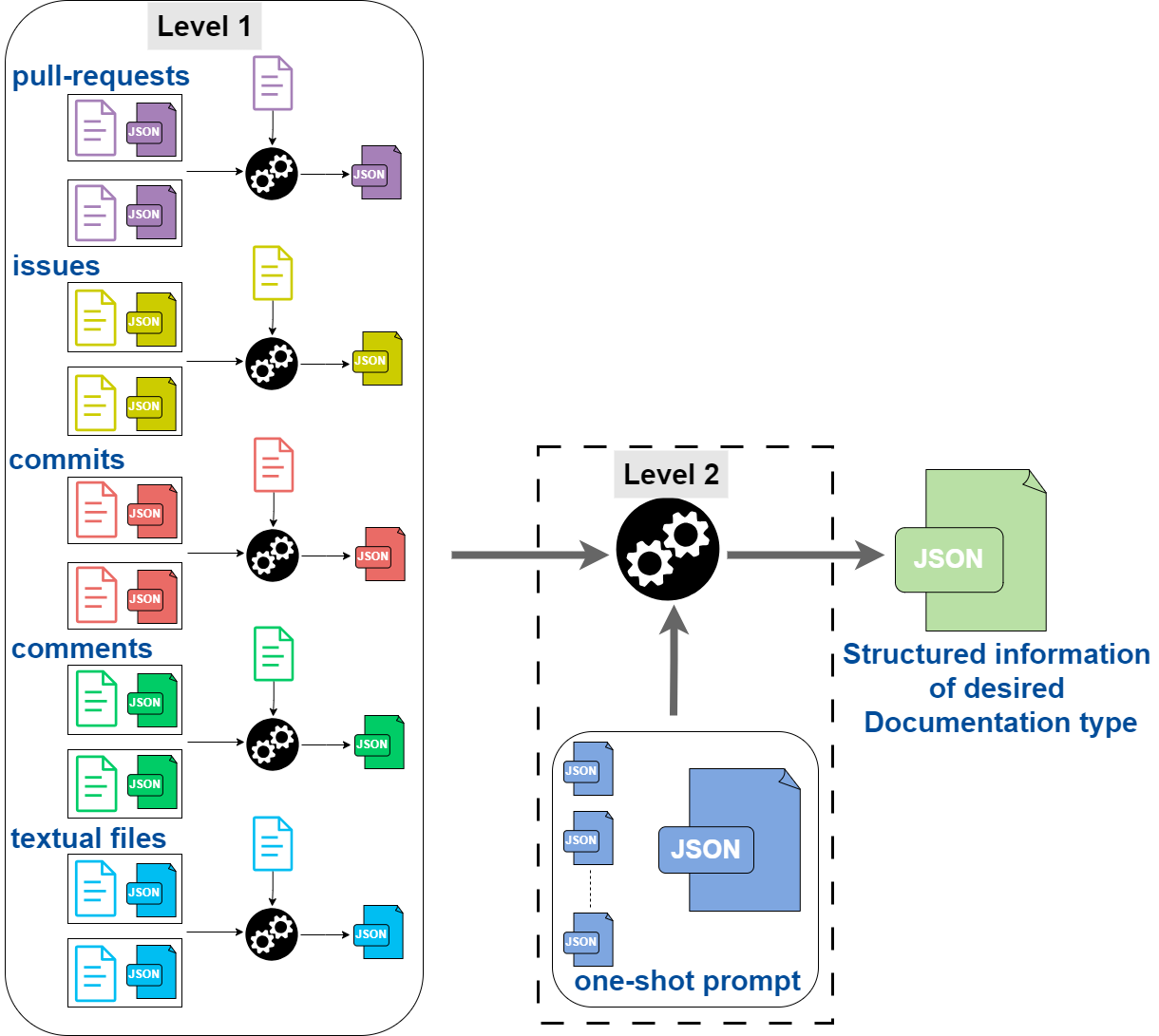}
    \caption{Complete architecture of \textit{DocFetch} to generate information corresponding to a documentation-type from five documentation sources.}
    \label{fig:arch}
\end{figure}
For each documentation type, the text extracted from a given source is paired with the corresponding \textit{intermediate} ground truth curated from that source. For example, for \textit{error-related} documentation, we use the \textit{intermediate} ground truth derived from \textit{pull-requests}, \textit{issues}, \textit{commits}, \textit{source code comments}, and \textit{textual files}, and input each pair into five LLM models in parallel, as shown in Fig.\ref{fig:arch}.

In one parallel pass, considering the \textit{pull-requests} as a documentation source, the text extracted from \textit{pull-requests} of the repository and the \textit{intermediate} ground-truth of \textit{error-related} documentation from \textit{pull-requests} are considered as the input-output pair. 
Two such input-output pairs are extracted from the \textit{intermediate} groundtruths corresponding to the desired documentation type, which we term as the two shots, as indicated in Fig. \ref{fig:level1}. We append the text extracted from \textit{pull-requests} for the desired repository to the prompt in the $<$context\_input$>$ of the prompt in Fig. \ref{fig:prompt}. Thus, in one parallel pass, two shots of input-output pairs corresponding to \textit{error-related} documentation from \textit{pull-requests} and the text extracted from \textit{pull-requests} of the desired repository are passed as a part of the prompt to the LLM model. Based on this prompt, the LLM model generates structured output corresponding to \textit{error-related} documentation from the \textit{pull-requests} of desired repository. Similar process is followed in the rest of the four parallel passes, and as a result, structured outputs corresponding to \textit{error-related} documentation from \textit{issues}, \textit{commits}, \textit{source code comments} and \textit{textual files}, along with \textit{pull-requests} are obtained in the level 1, which we consider as \textit{intermediate} outputs. 

The \textit{intermediate} outputs obtained in level 1 are passed as inputs to the level 2 as depicted in Fig. \ref{fig:arch}. The level 2 of the architecture comprises of one pre-trained LLM model that generates output based on the input prompt. We prompt the LLM model in level 2 to generate a structured information of the desired documentation type by consolidating the information in \textit{intermediate} outputs of level 1, which are passed as input in level 2 along with the prompt. To ensure that the output is in structured format and that the details in the \textit{intermediate} outputs are consolidated well, we pass an example of such \textit{intermediate} outputs and the corresponding \textit{final} output in the prompt. This example \textit{intermediate-final} output is the one-shot passed to the LLM model to fine-tune LLM towards ensuring in-context learning of the model. 

This example passed as one-shot corresponds to the \textit{intermediate} groundtruth and \textit{final} groundtruth pair of a random repository corresponding to the desired documentation type in the groundtruth dataset. The prompt for the level 2 LLM thus encapsulates a group of manually curated \textit{intermediate} groundtruths corresponding to the desired documentation type from all documentation sources of a random repository and the corresponding manually curated \textit{final} groundtruth of the desired documentation type. Similar to the prompts in level 1, the prompt for level 2 also expects the LLM to understand the analogy and generate the output in the structured format, by consolidating the context input (the \textit{intermediate} outputs). 

\section{Experimental Setup}

\label{exp}
In this section, we briefly discuss the usage of \textit{DocFetch}, with an example and further discuss the experimental setup followed to evaluate \textit{DocFetch}.

\begin{figure}[h!]
    \centering
    \includegraphics[width = \linewidth]{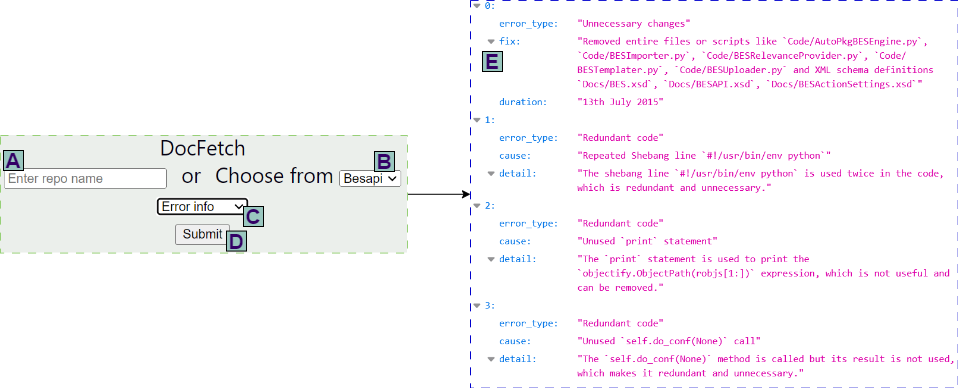}
    \caption{\textit{DocFetch} and an example of the generated output}
    \label{fig:user_scene}
\end{figure}

\textbf{User Scenario.}
\textit{DocFetch} facilitates the user to select a repository of choice to explore the documentation type information associated with that repository. The Fig. \ref{fig:user_scene} presents \textit{DocFetch} and the final generated output. The user can either enter the name of the repository of choice in the textbox indicated by Fig. \ref{fig:user_scene}[A] or to select the repository from a dropdown of repositories present in the \textit{DocMine} dataset from the dropdown indicated by Fig. \ref{fig:user_scene}[B]. The user can then choose the desired documentation type from the dropdown indicated by Fig. \ref{fig:user_scene}[C]. This dropdown consists all the five documentation types - \textit{API-related}, \textit{Error/Bug-related}, \textit{File-related}, \textit{License-related} and \textit{Project-related}. 
Upon choosing the repository of interest and the desired documentation type, the user can click on the \textit{Submit} button indicated by Fig. \ref{fig:user_scene}[D]. 

\textit{DocFetch} then processes the information extracted from the documentation sources corresponding to the repository of choice and with the help of LLM models in level 1 and level 2, it generates a final output corresponding to the desired documentation type in the `.json' format. This is available as a downloadable `.json' file for the user. For instance, let us assume that the user has selected \textit{Anvil} repository and expected \textit{error-related} information associated with that repository. The output of the \textit{DocFetch} tool in `.json' format corresponding to the \textit{error-related} documentation obtained from all the five documentation sources - \textit{pull-requests}, \textit{issues}, \textit{commits}, \textit{source code comments} and \textit{textual files} is presented in Fig. \ref{fig:user_scene}[E].

\textbf{Evaluation Process.}
We evaluate the performance of 
\textit{DocFetch} tool by answering the following research questions.
\begin{itemize}
    \item \textbf{RQ1:} How does \textit{DocFetch} perform while generating documentation type information from individual artifacts?
    
    \item \textbf{RQ2:} How does the performance of \textit{DocFetch} vary across different documentation types when all documentation sources are considered?
    
\end{itemize}

RQ1 focuses on analysing the performance of every LLM model in the level 1, while RQ2 analyses the performance of LLM model in level 2 of the \textit{DocFetch} architecture.
The performance of \textit{DocFetch} is derived by considering the results in both level 1 and level 2, as the generated output is dependant on both the levels. The \textit{intermediate} outputs generated in level 1 are passed as input to the model in level 2, thus indicating that the accuracy of results at the end of level 2 are dependant on the results of each parallel pass in level 1. This further indicates the need to analyse the individual performance of each model in level 1, which contributes to the overall performance of the tool, which is the focus of RQ1. Further, it is necessary to evaluate the performance of the LLM model in level 2, given that the output of the \textit{DocFetch} is directly dependant on the output at the end of level 2. The results of the performance of \textit{DocFetch} across different documentation types could indicate the reliability of information corresponding to a specific documentation type, thus motivating the need to answer RQ2.

\textit{\textbf{Baseline.}} We utilize the manually curated groundtruth dataset discussed in Section \ref{meth} as the baseline. We evaluate the performance of \textit{DocFetch} against this baseline dataset. As discussed in Section \ref{meth}, the groundtruth dataset contains information corresponding to five different documentation types extracted from five different documentation sources. This accounts to 450 \textit{intermediate} groundtruths and 90 \textit{final} groundtruths across 18 repositories, obtained from a significant sample (1205 artifacts) of 1.35M artifacts in \textit{DocMine}. 

\textit{\textbf{Dataset.}} For the purpose of experimental analysis, we generate the \textit{intermediate} outputs from level 1 of the \textit{DocFetch} architecture corresponding to the 18 repositories considered in the groundtruth dataset. Thus, using the prompt discussed earlier in Fig. \ref{fig:prompt}, we generate structured information for each of the five documentation types based on the content present in the text extracted from \textit{pull-requests}, \textit{issues}, \textit{commits}, \textit{source code comments} and \textit{textual files} individually. Thus, we automatically analyse each documentation source corresponding to each of the 18 repositories and generate 450 \textit{intermediate} outputs. Using the corresponding five \textit{intermediate} outputs for every documentation type, for a given repository, we generate the \textit{final} structured information consolidated from all documentation sources using level 2 of the \textit{DocFetch} architecture. This process finally results in 90 \textit{final} outputs corresponding to the 18 repositories, which can be evaluated against the 90 \textit{final} groundtruths. 

\textit{\textbf{Evaluation Metrics.}} To evaluate the performance of \textit{DocFetch}, towards answering RQ1 and RQ2, we compare the generated dataset with the baseline against  BLEU and ROUGE scores. These metrics  are widely used in the literature to evaluate the performance of text generation models based on similarity \cite{chang2024survey}. BLEU score and ROUGE scores were used to evaluate the performance of GPT-3.5 and other pre-trained LLMs involved in various software documentation generation tasks such as generation of source code comments \cite{khan2022automatic} and code summaries \cite{ahmed2024automatic, ahmed2022few}. 



Inline with the literature, and considering the relevance of BLEU and ROUGE metrics in comparing the similarity and quality of machine generated text with manually curated text, we also employ BLEU-4 and ROUGE-L metrics to evaluate the performance of \textit{DocFetch}.

\begin{tcolorbox}[  colback=gray!5!white,colframe=gray!75!black,
  fonttitle=\bfseries, title= ]
\textit{DocFetch} is evaluated using \textit{DocMine} dataset, by considering the  groundtruth created in Section \ref{meth} as baseline, against the BLEU and ROUGE scores as metrics for evaluation.
\end{tcolorbox}

\section{Results}
\label{res}
In this section, we discuss the results obtained after evaluation of \textit{DocFetch}, along with a few examples, and answer the research questions discussed in the previous section.

\textbf{Performance of \textit{DocFetch} in generating documentation type information from individual artifacts (RQ1).}

We generate the documentation type information from each of the documentation sources and calculate the BLEU-4 and ROUGE-L scores for the respective generated information.

Table \ref{tab:rr} presents the ROUGE-L scores for generating each documentation type from the documentation sources. Of all, 
the license-related information generated from \textit{pull-requests} had the highest ROUGE-L score of 0.71, followed by 
the error-related information generated from  \textit{commits} with a ROUGE-L score of 0.6. Further, \textit{commits} had the ROUGE-L scores of 0.52, 0.55, 0.35 and 0.44 for api-related, file-related, license-related and project-related information generated from them respectively. 

\begin{table}[h!]
\centering
\small\setlength\tabcolsep{2.2pt}
\footnotesize
\caption{Performance of \textit{DocFetch} to Generate Documentation Type Information from Each of the Documentation Sources Based on ROUGE-L Scores}
\label{tab:rr}
\begin{tabular}{|l|c|c|c|c|c|}
\hline
                         & \textbf{Pull-Requests} & \textbf{Issues} & \textbf{Commits} & \textbf{Comments} & \textbf{TextualFiles} \\ \hline
\textbf{API-related}     & 0.54                   & \textbf{0.57}             & 0.52              & 0.43               & 0.39                  \\ \hline
\textbf{Error-related}   & 0.48                  & 0.37            & \textbf{0.6}             & 0.35              & 0.35                  \\ \hline
\textbf{File-related}    & 0.55                   & 0.42            & 0.55             & \textbf{0.56}              & 0.2                   \\ \hline
\textbf{License-related} & \textbf{0.71 }                   & 0.32            & 0.35             & 0.66              & 0.42                  \\ \hline
\textbf{Project-related} & \textbf{0.49  }                 & 0.18            & 0.44             & 0.36               & 0.3                   \\ \hline
\end{tabular}
\end{table}

Following \textit{commits}, \textit{pull-requests} exhibited next higher ROUGE-L scores for generation of information corresponding to each of the documentation sources. The ROUGE-L scores for information generated from \textit{pull-requests} were 0.54, 0.48, 0.55, and 0.49 for \textit{api}, \textit{error}, \textit{file} and \textit{project} related information respectively. The ROUGE-L scores 
for \textit{api} and \textit{error} related information extracted from \textit{commits} and \textit{issues} is greater than majority of the ROUGE-L scores achieved in state-of-the-art approaches that use pre-trained LLMs to generate source code summaries \cite{ahmed2024automatic} (which are commonly considered as documentation of the projects).

\begin{figure}[h!]
    \centering
    \includegraphics[scale=0.5]{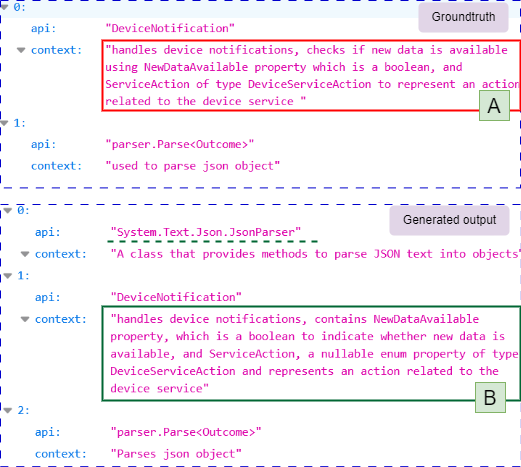}
    \caption{Example of API-related information generated by \textit{DocFetch} from \textit{issues}, corresponding to \textit{FastJsonParser} repository and its corresponding groundtruth.}
    \label{fig:api-issue-example}
    \vspace{-0.5cm}
\end{figure}
\begin{figure}[h!]
    \centering    \includegraphics[width=0.7\linewidth]{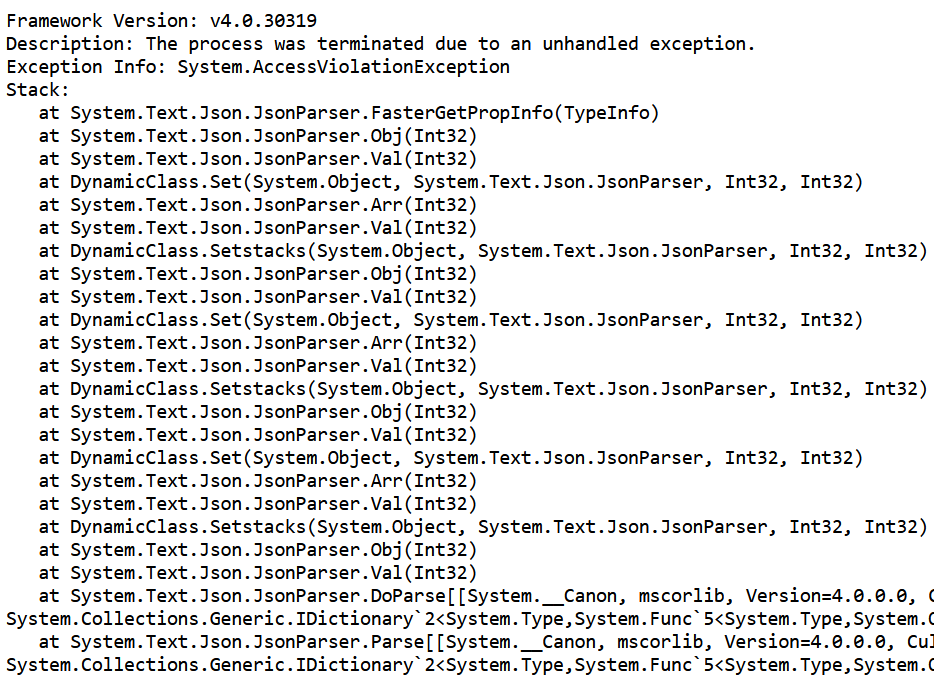}
    \caption{A snipped block of \textit{issues} extracted from \textit{FastJsonParser} repository, depicting the occurrence of reference to \textit{System.Text.Json.JsonParser} library.}
    \label{fig:apiexample}
\end{figure}
 Fig. \ref{fig:api-issue-example} represents \textit{api-related} information generated by \textit{DocFetch} from \textit{issues} extracted from the \textit{FastJsonParser} repository (in the \textit{DocMine} dataset), and the manually curated ground-truth for the same.
 The information in the context corresponding to \textit{DeviceNotification} API, generated by \textit{DocFetch}, highlighted in Fig. \ref{fig:api-issue-example} [B] mentions the \textit{enum} data type of \textit{ServiceAction} property, the detail of which is not included in the groundtruth dataset (indicated in Fig. \ref{fig:api-issue-example} [A]. Also, the generated output comprises information about \textit{System.Text.Json.JsonParser} library (highlighted by a dashed line in Fig. \ref{fig:api-issue-example}), which is not present in the corresponding groundtruth. 

The \textit{System.Text.Json.JsonParser} library is indicated in the \textit{stack trace} of the \textit{issues} extracted, to represent a violation, while an \textit{issue} is being logged, with its references being those indicated in Fig. \ref{fig:apiexample}.

Since there is no specific discussion around the \textit{System.Text.Json.JsonParser} library, this library was not included in the groundtruth creation. However, \textit{DocFetch} has absorbed the context in which this library is being used, and presented the functionality of this library. This example indicates an instance where the generated output has deeper levels of detail than the groundtruth.

Documentation types generated from \textit{comments} and \textit{textual files} generally had the lowest ROUGE-L scores compared to other sources, except for \textit{license} and \textit{file}-related information. \textit{License-related} information from \textit{comments} scored 0.66, outperforming \textit{commits} (0.35), \textit{issues} (0.32), and \textit{textual files} (0.42). Similarly, \textit{file-related } information from \textit{comments} (0.56) exceeded that from \textit{textual files} (0.20). For other types, \textit{comments} scored lower: 0.35 (error), 0.43 (API), 0.56 (pull-requests), and 0.36 (project). Notably, project-related data from \textit{issues} had the lowest score overall at 0.18. The lowest ROUGE-L scores were mostly from \textit{textual files}, with 0.39, 0.35, 0.20, 0.42, and 0.30 for API, error, file, license, and project documentation respectively.
\begin{table}[h!]
\centering
\small\setlength\tabcolsep{2.2pt}
\footnotesize
\caption{Performance of \textit{DocFetch} to Generate Documentation Type Information from Each of the Documentation Sources Based on BLEU Scores}
\label{tab:blue}
\begin{tabular}{|l|c|c|c|c|c|}
\hline
                         & \textbf{Pull-Requests} & \textbf{Issues} & \textbf{Commits} & \textbf{Comments} & \textbf{TextualFiles} \\ \hline
\textbf{API-related}     & 35.13\%                   & 34.28\%          & \textbf{35.65\% }         & 28.98\%            & 28.11\%               \\ \hline
\textbf{Error-related}   & 31.21\%                & 33.43\%          &\textbf{ 39.63\%  }         & 26.19\%           & 29.14\%               \\ \hline
\textbf{File-related}    & 27.21\%                & 30.2\%         & \textbf{36.42\%}          & 24.25\%            & 22.93\%               \\ \hline
\textbf{License-related} & 35.05\%                & 26.1\%         & 25.19\%          & \textbf{32.88\% }          & 30.82\%               \\ \hline
\textbf{Project-related} & 28.3\%                & 21.18\%         & \textbf{41.47\%}          & 25.2\%            & 24.6\%                \\ \hline
\end{tabular}

\end{table}



We observe that the BLEU-4 scores for all the documentation sources is higher while generating \textit{error-related} and \textit{api-related} documentation than the other documentation types. Further, the BLEU-4 scores for commits and issues were observed to be higher than the other artifacts, indicating that the documentation-type related information could be well consolidated from these two sources than pull-requests, textual files and source code comments. The BLEU-4 for majority of the information generated are close to 30\%, which indicates `good' performance of the approach \cite{ahmed2024automatic}. The BLEU-4 scores for \textit{project} and \textit{error-related} information generated from \textit{commits} are the highest (41.47\% and 39.63\%), which are greater than majority of the BLEU scores achieved in the state-of-the-art code summarization tasks \cite{ahmed2024automatic, khan2022automatic}.  The BLEU-4 and ROUGE-L scores indicate that the generated information is inline with the groundtruths for majority of the documentation sources. This indicates that the pre-trained LLM models in the level-1 have good performance with two-shot prompts.

\begin{tcolorbox}[  colback=gray!5!white,colframe=gray!75!black,
  fonttitle=\bfseries, title= Answering RQ1]
\textit{DocFetch} showed good performance while generating \textit{error-related} information from \textit{commits}, with ROUGE-L score of 0.6, while it exhibited limited performance for generating \textit{project-related} information from \textit{issues} and \textit{file-related} information from \textit{textual files}.
\end{tcolorbox}

\textbf{Performance of \textit{DocFetch} in generating different documentation types acorss all the documentation sources (RQ2).}
Similar to answering RQ1, we calculate the BLEU-4 and ROUGE-L scores for the information corresponding to each documentation type generated by consolidating the information in \textit{intermediate} outputs generated from each documentation source. The results of the analysis are presented in Table \ref{tab:all}. Of all documentation types, information extracted for \textit{file}-related documentation type had the highest ROUGE-L score of 0.39, followed by \textit{license-related} documentation type, with ROUGE-L score of 0.33. \textit{Project-related} documentation type had the lowest ROUGE-L score of 0.24.  The BLEU-4 scores were 36.87\%, 43.24\%, 41.37\%, 28.84\% and 28.9\% for \textit{error}, \textit{api}, \textit{file}, \textit{license} and \textit{project} related information generated respectively.

\begin{table}[h!]
\centering
\footnotesize
\caption{Performance of \textit{DocFetch} by Integrating Content across All the Documentation Sources Based on ROUGE-L and BLEU Scores}
\label{tab:all}
\begin{tabular}{|l|c|c|c|}
\hline
                         & \textbf{ROUGE-L} & \textbf{BLEU-4} \\ \hline
\textbf{API-related}     & 0.28                      & \textbf{43.24\% }        \\ \hline
\textbf{Error-related}   & 0.28   & 36.87\%         \\ \hline
\textbf{File-related}    & 0.39       & \textbf{41.37\%}         \\ \hline
\textbf{License-related} & 0.33        & 28.84\%         \\ \hline
\textbf{Project-related} & 0.24          & 28.9\%         \\ \hline
\end{tabular}

\end{table}

While BLEU-4 scores ($>=$30\%) suggest good translation quality, we manually reviewed 50 randomly selected outputs, with ten outputs obtained from each documentation source (two each for every documentation type). We manually compared these outputs generated with groundtruth dataset, to understand areas where the model could be improved. This exploration revealed that, despite the scope for improvement in the scores, the model performed better, with more information presented in the generated data than that which is present in the groundtruth dataset. This indicates the capability of the tool to generate data that is deducable beyond manual curation. Our manual analysis revealed about 3\% of the instances where the LLM did not produce the output in the `.json' format, with preceding and succeeding textual sentences encompassing the desired `.json' format. Although the number of instances is low, in comparison to the outputs with expected formats, such instances can be avoided by further processing. 

\begin{tcolorbox}[  colback=gray!5!white,colframe=gray!75!black,
  fonttitle=\bfseries, title= Answering RQ2]
\textit{DocFetch} exhibited good performance while generating \textit{file-related} and \textit{api-related} information with higher ROUGE and BLEU scores for these documentation types.
\end{tcolorbox}

\section{Threats to Validity}
\label{ttv}
We discuss the threats to validity associated with \textit{DocFetch} below.

\textbf{Internal Validity-}
The groundtruth dataset created through manual analysis of the text extracted from the documentation sources might be biased towards the opinion of the author. To avoid such bias, we validated the groundtruth created with two other researchers and made necessary modifications after thorough discussions. 
During few-shot training, we take care to steer the LLM toward repository-specific generation via in-context learning. However, due to the influence of the model's pretraining corpus, instances of out-of-context or hallucinated information, not specific to a repository may still occur.


Further, there could be bias creeping in the performance of LLM due to the repositories used as few-shot prompts. The LLM might completely rely on the context of the prompts including the content being passed, while the intention of few-shot prompts is only to provide context in terms of the format in which the output should bbe generated. However, we avoid the possibility of bias that might be induced in the results by using same repository information as few-shots, by altering the few-shot repositories for every pass. Further, we employ BLEU and ROUGE scores to evaluate the performance of \textit{DocFetch} which are the metrics commonly used to evaluate the performance of approaches that use LLMs to ensure an accurate interpretation of the results.

\textbf{Construct Validity-}
We attempt to pass the entire textual data extracted from the artifacts. However, usage of any neural network models, similar to LLMs restricts usage of large files due to the limited context window size. While we can use the sliding window approach of breaking the large files into chunks and processing them in steps, this is time consuming and results in increased wait times for the end users. Also, in spite of passing chunks of the files as input, the outputs for every chunk need to be consolidated towards obtaining the \textit{intermediate} output corresponding to a specific file, which in itself could be a large file. This \textit{intermediate} output is passed as input to the next level LLM along with other similar \textit{intermediate} outputs. 

As a result, if the \textit{intermediate} output of one file alone is large, then group of such \textit{intermediate} outputs would again result in exceeding context window size. Chunking these \textit{intermediate} outputs further compromises the accuracy of the results. Hence, we restrict the sizes of the input files being processed. While we do not discard large files, we process only a part of them to remain within the context window limits. However, the \texttt{GFLP2.0} LLM used for building \textit{DocFetch} supports context window size of 37K characters, which greater than other LLMs such as \textit{gpt-turbo}, which supports a context-window of about 4K characters.

\textbf{External Validity-} 
We use \texttt{GFLP2.0} as the LLM model in both the levels of \textit{DocFetch} approach. While the approach remains unvariable, the results generated by the LLM are not always identical, and keep varying. However, we observe that the context of the generated information is mostly retained, irrespective of the varying sentences and their formation. Such varying results could alter the BLEU and ROUGE scores upon regeneration of the information. 

Further, the BLEU and ROUGE scores are also dependant on the curation of groundtruths, which are in turn dependant on the author's perspective and might vary from researcher to researcher. Re-curating the groundtruths, would thus also effect the BLEU and ROUGE scores of \textit{DocFetch}. The results obtained for each documentation type are also specific to the \texttt{GFLP2.0} model. Replacing the \texttt{GFLP2.0} model with another LLM might yield different set of results for the documentation types. Finally, the \textit{DocFetch} currently supports only the repositories in the \textit{DocMine} dataset. While this can be extended to repositories outside the \textit{DocMine} dataset, the time taken to obtain results for such repositories will be longer, as the data corresponding to all documentation sources of these repositories has to be scraped on the go using the GitHub API. Such scraping is dependant on the network speed and the GitHub API limits, unlike the offline data available for \textit{DocMine} repositories.

\section{Related Work}
\label{rw}
In this section we discuss the some of the works in the literature that employed large language models for various downstream software engineering tasks. Table \ref{tab:relwork} presents a brief overview of the works discussed in this section.

\begin{table}[!t]
\centering
\footnotesize
\renewcommand\cellalign{tl}
\renewcommand\theadalign{tl}
\begin{tabularx}{\columnwidth}{|c|X|X|}
\hline
\textbf{Task} & \textbf{LLM Used} & \textbf{Dataset Used} \\ \hline
Code Generation \cite{mathews2024test} & GPT-4, Llama3 & MBPP, HumanEval, CodeChef\\ \hline
Code Generation \cite{ren2023misuse} & ChatGPT with API and exception knowledge & 3,079 Java code generation tasks \\ \hline
Test Case Generation \cite{wang2024hits} & \texttt{gpt-3.5-turbo} & Complex methods from 10 Java projects \\ \hline
Test Case Generation \cite{schafer2023empirical} & \texttt{gpt-3.5-turbo} with API info & 1,684 JavaScript API functions \\ \hline
Test Case Generation \cite{lemieux2023codamosa} & Codex & 486 benchmarks \\ \hline
Test Case Generation \cite{dakhel2024effective} & \makecell[l]{\texttt{code-davinci}\\\texttt{-002}, \texttt{llama-2-}\\\texttt{chat}, mutants} & HumanEval and Refactory \\ \hline
Commit Messages \cite{wu2025empirical} &\texttt{gpt-3.5-turbo}, DeepSeek-V2-Chat,
Claude-3-Haiku & MCMD \\ \hline
Code Summarization \cite{khan2022automatic} & Codex (zero-shot, one-shot) & CodeSearchNet \\ \hline
Code Summarization \cite{ahmed2022few} & Codex (few-shot: in-project, cross-project) & CodeSearchNet \\ \hline
Code Summarization \cite{ahmed2024automatic} & \texttt{gpt-3.5-turbo} with semantic prompts & CodeSearchNet \\ \hline
Code Summarization \cite{su2024distilled} & Distilled GPT-3.5 & \makecell[l]{\texttt{funcom-}\\\texttt{java-long}} \\ \hline
\end{tabularx}
\caption{Studies using pre-trained LLMs for downstream software engineering tasks.}
\label{tab:relwork}
\end{table}

\subsection{Using LLMs to Generate Software Artifacts}
With the growing popularity and performance of pre-trained LLMs, researchers have proposed various approaches using these LLMs to address multiple challenges in software engineering. LLMs are thus used for multiple downstream tasks ranging from code generation to test case generation to program repair.

Mathews et al. proposed inclusion of testcases along with the problem statements to generate more accurate python code using GPT-4 and Llama 3 models \cite{mathews2024test}. They evaluated this approach on MBPP, HumanEval and CodeChef datasets and observed an improvement of about 38.6\%.
Ren et al. employed an integrated approach using ChatGPT to generate java source code with improved exception handling \cite{ren2023misuse}. They performed knowledge driven prompting to ChatGPT to obtain information about APIs and exception handling from knowledge extracted from API documentation, which is further used to generate the source code with better exception handling \cite{ren2023misuse}.

Lemieux et al. proposed CODAMOSA, an approach to generate testcases during search based software testing (SBST) when a stall occurs in the coverage using pre-trained LLM, Codex \cite{lemieux2023codamosa}. They observed high convergence in the generated testcases than the state-of-the-art approaches over 486 benchmarks.
To further improve the bug detection capabilities of the generated testcases, Dakhel et al. \cite{dakhel2024effective} presented an integrated approach using mutation testing and the LLM models - \texttt{code-davinci-002} and \texttt{llama-2-chat}. They augmented the prompts to LLMs with initial test case versions and the surviving mutants associated with code to be tested. This approach showed an increase of 28\% in detecting faulty code snippets in comparison with the state-of-the-art models. 

Test cases were also generated using \texttt{gpt-3.5-turbo} with API functions and associated comments and usage snippets as prompts \cite{schafer2023empirical}. These generated testcases revealed about 70\% statement coverage and 52.8\% branch coverage over 1,684 Javascript API functions. Unit test cases were generated by prompting \texttt{gpt-3.5-turbo} to decompose complex Java methods into logical slices \cite{wang2024hits}. This resulted in an average branch coverage of about 48\% and line coverage of 55\%.


\subsection{Using LLMs for Software documentation}
Along with multiple downstream software engineering tasks, LLMs have been employed to generate software documentation corresponding to source code summaries. 

Commit messages were generated using in-context learning where code-diffs and corresponding messages were passed as prompts to the LLMs \cite{wu2025empirical}. This approach has been compared with various LLM models such as GPT-3.5-Turbo, DeepSeek-V2-Chat, Claude-3-Haiku and so on, against MCMD datasets, where GPT-3.5-Turbo and DeepSeek-V2-Chat were observed to be performing better in both subjective evaluations and on new datasets. 

GPT-based Codex model has been used to generate one-line method summaries using one-shot and zero-shot prompts \cite{khan2022automatic}. This model has been evaluated against the CodeSearchNet \cite{husain2019codesearchnet} dataset and was observed to perform better than the state-of-the-art models with an average BLEU score of 20.6 against code snippets in six different programming languages. In another instance, Codex model has been used to generate source code summaries based on few-shot prompts corresponding to cross-project and in-project examples \cite{ahmed2022few}. Ten examples of source code and corresponding explanations belonging to different projects and the same project were passed as prompts in different passes, where the summaries generated through few-shots from cross-project yielded BLEU-4 score of 21.65\% while the summaries generated from in-project few shots yielded BLEU-4 score of 24.37\%, indicating appreciable performance of the model \cite{ahmed2022few}.

Ahmed et al. further attempted to augment the prompts being passed to LLMs with the semantic information to generate better code summaries \cite{ahmed2024automatic}. They extracted semantic information corresponding to the dataflow, scope and names of the variables associated with the code snippet using Treesitter and passed the same as prompt to the Codex model along with source code and repository information to generate code summaries. This approach is also evaluated against CodeSearchNet dataset which yielded BLEU score of 32.73\% which is relatively larger the then existing state-of-the-art models for code snippets in PHP language \cite{ahmed2024automatic}. 

Further, researchers also proposed an in-house large language model to generate code summaries without passing the code to untrusted third parties \cite{su2024distilled}. This in-house large language model is a distilled GPT-3.5 that can be  run on a 16GB GPU machine, trained using the code summaries generated by GPT-3.5 to enable the model to be a small scale model that would perform similar to GPT-3.5 on the tasks it is trained for (code sumamrization). They trained the decoder only transformer \texttt{jam} and \texttt{starcoder} with 2.15 million java methods and associated summaries generated using GPT-3.5 and obtained a METEOR score of 40.73 for 350m parameter \texttt{jam} and 44.8 for \texttt{starcoder} models \cite{su2024distilled}.

\section{Conclusion and Future Work}
\label{concl}
We present \textit{DocFetch}, a tool to generate documentation type information from various software artifacts associated with the repositories. We discuss the approach of the tool where we employed pre-trained LLM, \texttt{GFLP2.0}, at two levels, which accounted to a total of 6 \textit{GFLP2.0} instances. These LLMs were provided with the context of the desired output format and structure through few-shot and one-shot prompts at levels 1 and 2 respectively. \textit{DocFetch} tool considers documentation types and sources discussed in \cite{venigalla2024exploratory} and uses the \textit{DocMine} \cite{venigalla2023docmine} dataset to generate documentation type information in `.json' structured format, thus transforming the unstructured textual information. 

We further conducted experiments to assess the performance of the tool by comparing the results generated by \textit{DocFetch} against manually curated groundtruth dataset. We observe that \textit{DocFetch} performs well across generation of all the documentation types, with an average BLEU-4 score of 50\%. Manual analysis of the generated results further indicated good performance of \textit{DocFetch}, with the generated text being more informative than the manually curated text in several instances, in spite of low BLEU and ROUGE scores. These performance scores demonstrated the capability of 
\textit{DocFetch} to generate different documentation type information, thus providing on-demand documentation for a repository based on the user requirement. 

As a part of the future work, we plan to extend \textit{DocFetch} to support more number of repositories, by optimizing the dynamic scraping of information, thus extending the scope of \textit{DocFetch}, beyond the \textit{DocMine} dataset. 
Further, we plan to facilitate users to enter a date range between which the users would require information about the project. This can also help us in optimizing and regulating the amount of information required to be scraped, thus also facilitating support to repositories beyond the \textit{DocMine} dataset. Also, we plan to expand the groundtruth dataset and evaluate it with larger group of researchers and practitioners to ensure the quality and unbiased nature of the groundtruth dataset.

\balance
\bibliography{references}

\begin{thebibliography}{10}
\providecommand{\url}[1]{#1}
\csname url@samestyle\endcsname
\providecommand{\newblock}{\relax}
\providecommand{\bibinfo}[2]{#2}
\providecommand{\BIBentrySTDinterwordspacing}{\spaceskip=0pt\relax}
\providecommand{\BIBentryALTinterwordstretchfactor}{4}
\providecommand{\BIBentryALTinterwordspacing}{\spaceskip=\fontdimen2\font plus
\BIBentryALTinterwordstretchfactor\fontdimen3\font minus \fontdimen4\font\relax}
\providecommand{\BIBforeignlanguage}[2]{{%
\expandafter\ifx\csname l@#1\endcsname\relax
\typeout{** WARNING: IEEEtran.bst: No hyphenation pattern has been}%
\typeout{** loaded for the language `#1'. Using the pattern for}%
\typeout{** the default language instead.}%
\else
\language=\csname l@#1\endcsname
\fi
#2}}
\providecommand{\BIBdecl}{\relax}
\BIBdecl

\bibitem{garousi2015usage}
G.~Garousi, V.~Garousi-Yusifo{\u{g}}lu, G.~Ruhe, J.~Zhi, M.~Moussavi, and B.~Smith, ``Usage and usefulness of technical software documentation: An industrial case study,'' \emph{Information and Software Technology}, vol.~57, pp. 664--682, 2015.

\bibitem{alamin2023developer}
M.~A.~A. Alamin, G.~Uddin, S.~Malakar, S.~Afroz, T.~Haider, and A.~Iqbal, ``Developer discussion topics on the adoption and barriers of low code software development platforms,'' \emph{Empirical software engineering}, vol.~28, no.~1, p.~4, 2023.

\bibitem{borges2018s}
H.~Borges and M.~T. Valente, ``What’s in a github star? understanding repository starring practices in a social coding platform,'' \emph{Journal of Systems and Software}, vol. 146, pp. 112--129, 2018.

\bibitem{gao2025adapting}
H.~Gao, C.~Treude, and M.~Zahedi, ``Adapting installation instructions in rapidly evolving software ecosystems,'' \emph{IEEE Transactions on Software Engineering}, 2025.

\bibitem{liu2021reproducibility}
C.~Liu, C.~Gao, X.~Xia, D.~Lo, J.~Grundy, and X.~Yang, ``On the reproducibility and replicability of deep learning in software engineering,'' \emph{ACM Transactions on Software Engineering and Methodology (TOSEM)}, vol.~31, no.~1, pp. 1--46, 2021.

\bibitem{mitchell2022}
\BIBentryALTinterwordspacing
M.~Joblin and S.~Apel, ``How do successful and failed projects differ? a socio-technical analysis,'' \emph{ACM Trans. Softw. Eng. Methodol.}, dec 2021. [Online]. Available: \url{https://doi.org/10.1145/3504003}
\BIBentrySTDinterwordspacing

\bibitem{ehsan2020empirical}
O.~Ehsan, S.~Hassan, M.~E. Mezouar, and Y.~Zou, ``An empirical study of developer discussions in the gitter platform,'' \emph{ACM Transactions on Software Engineering and Methodology (TOSEM)}, vol.~30, no.~1, pp. 1--39, 2020.

\bibitem{xiao2023early}
W.~Xiao, H.~He, W.~Xu, Y.~Zhang, and M.~Zhou, ``How early participation determines long-term sustained activity in github projects?'' in \emph{Proceedings of the 31st ACM Joint European Software Engineering Conference and Symposium on the Foundations of Software Engineering}, 2023, pp. 29--41.

\bibitem{bao2019large}
L.~Bao, X.~Xia, D.~Lo, and G.~C. Murphy, ``A large scale study of long-time contributor prediction for github projects,'' \emph{IEEE Transactions on Software Engineering}, vol.~47, no.~6, pp. 1277--1298, 2019.

\bibitem{venigalla2025there}
A.~S.~M. Venigalla and S.~Chimalakonda, ``Is there a correlation between readme content and project meta-characteristics?'' \emph{Software: Practice and Experience}, vol.~55, no.~3, pp. 589--609, 2025.

\bibitem{aggarwal2014co}
K.~Aggarwal, A.~Hindle, and E.~Stroulia, ``Co-evolution of project documentation and popularity within github,'' in \emph{Proceedings of the 11th Working Conference on Mining Software Repositories}, 2014, pp. 360--363.

\bibitem{wang2023study}
T.~Wang, S.~Wang, and T.-H.~P. Chen, ``Study the correlation between the readme file of github projects and their popularity,'' \emph{Journal of Systems and Software}, vol. 205, p. 111806, 2023.

\bibitem{fan2021makes}
Y.~Fan, X.~Xia, D.~Lo, A.~E. Hassan, and S.~Li, ``What makes a popular academic ai repository?'' \emph{Empirical Software Engineering}, vol.~26, no.~1, pp. 1--35, 2021.

\bibitem{aghajani2019software}
E.~Aghajani, C.~Nagy, O.~L. Vega-M{\'a}rquez, M.~Linares-V{\'a}squez, L.~Moreno, G.~Bavota, and M.~Lanza, ``Software documentation issues unveiled,'' in \emph{2019 IEEE/ACM 41st International Conference on Software Engineering (ICSE)}.\hskip 1em plus 0.5em minus 0.4em\relax IEEE, 2019, pp. 1199--1210.

\bibitem{wu2024comprehensive}
J.~Wu, H.~He, K.~Gao, W.~Xiao, J.~Li, and M.~Zhou, ``A comprehensive analysis of challenges and strategies for software release notes on github,'' \emph{Empirical Software Engineering}, vol.~29, no.~5, p. 104, 2024.

\bibitem{rong2024code}
G.~Rong, Y.~Yu, S.~Liu, X.~Tan, T.~Zhang, H.~Shen, and J.~Hu, ``Code comment inconsistency detection and rectification using a large language model,'' in \emph{2025 IEEE/ACM 47th International Conference on Software Engineering (ICSE)}.\hskip 1em plus 0.5em minus 0.4em\relax IEEE Computer Society, 2024, pp. 432--443.

\bibitem{huang2024empirical}
Z.~Huang, Y.~Huang, X.~Chen, X.~Zhou, C.~Yang, and Z.~Zheng, ``An empirical study on learning-based techniques for explicit and implicit commit messages generation,'' in \emph{Proceedings of the 39th IEEE/ACM International Conference on Automated Software Engineering}, 2024, pp. 544--556.

\bibitem{liu2020atom}
S.~Liu, C.~Gao, S.~Chen, L.~Y. Nie, and Y.~Liu, ``Atom: Commit message generation based on abstract syntax tree and hybrid ranking,'' \emph{IEEE Transactions on Software Engineering}, vol.~48, no.~5, pp. 1800--1817, 2020.

\bibitem{wei2020retrieve}
B.~Wei, Y.~Li, G.~Li, X.~Xia, and Z.~Jin, ``Retrieve and refine: exemplar-based neural comment generation,'' in \emph{Proceedings of the 35th IEEE/ACM International Conference on Automated Software Engineering}, 2020, pp. 349--360.

\bibitem{bansal2023function}
A.~Bansal, Z.~Eberhart, Z.~Karas, Y.~Huang, and C.~McMillan, ``Function call graph context encoding for neural source code summarization,'' \emph{IEEE Transactions on Software Engineering}, 2023.

\bibitem{gao2023code}
S.~Gao, C.~Gao, Y.~He, J.~Zeng, L.~Nie, X.~Xia, and M.~Lyu, ``Code structure--guided transformer for source code summarization,'' \emph{ACM Transactions on Software Engineering and Methodology}, vol.~32, no.~1, pp. 1--32, 2023.

\bibitem{eliseeva2023commit}
A.~Eliseeva, Y.~Sokolov, E.~Bogomolov, Y.~Golubev, D.~Dig, and T.~Bryksin, ``From commit message generation to history-aware commit message completion,'' in \emph{2023 38th IEEE/ACM International Conference on Automated Software Engineering (ASE)}.\hskip 1em plus 0.5em minus 0.4em\relax IEEE, 2023, pp. 723--735.

\bibitem{li2024understanding}
C.~Li, Z.~Xu, P.~Di, D.~Wang, Z.~Li, and Q.~Zheng, ``Understanding code changes practically with small-scale language models,'' in \emph{Proceedings of the 39th IEEE/ACM International Conference on Automated Software Engineering}, 2024, pp. 216--228.

\bibitem{khan2022automatic}
J.~Y. Khan and G.~Uddin, ``Automatic code documentation generation using gpt-3,'' in \emph{Proceedings of the 37th IEEE/ACM International Conference on Automated Software Engineering}, 2022, pp. 1--6.

\bibitem{nam2024using}
D.~Nam, A.~Macvean, V.~Hellendoorn, B.~Vasilescu, and B.~Myers, ``Using an llm to help with code understanding,'' in \emph{Proceedings of the IEEE/ACM 46th International Conference on Software Engineering}, 2024, pp. 1--13.

\bibitem{jamieson2024predicting}
J.~Jamieson, N.~Yamashita, and E.~Foong, ``Predicting open source contributor turnover from value-related discussions: An analysis of github issues,'' in \emph{Proceedings of the 46th IEEE/ACM International Conference on Software Engineering}, 2024, pp. 1--13.

\bibitem{kuramoto2022visual}
H.~Kuramoto, M.~Kondo, Y.~Kashiwa, Y.~Ishimoto, K.~Shindo, Y.~Kamei, and N.~Ubayashi, ``Do visual issue reports help developers fix bugs? a preliminary study of using videos and images to report issues on github,'' in \emph{Proceedings of the 30th IEEE/ACM International Conference on Program Comprehension}, 2022, pp. 511--515.

\bibitem{xiao2022recommending}
W.~Xiao, H.~He, W.~Xu, X.~Tan, J.~Dong, and M.~Zhou, ``Recommending good first issues in github oss projects,'' in \emph{Proceedings of the 44th International Conference on Software Engineering}, 2022, pp. 1830--1842.

\bibitem{mazuera2022taxonomy}
A.~Mazuera-Rozo, C.~Escobar-Vel{\'a}squez, J.~Espitia-Acero, D.~Vega-Guzm{\'a}n, C.~Trubiani, M.~Linares-V{\'a}squez, and G.~Bavota, ``Taxonomy of security weaknesses in java and kotlin android apps,'' \emph{Journal of Systems and Software}, vol. 187, p. 111233, 2022.

\bibitem{golzadeh2021ground}
M.~Golzadeh, A.~Decan, D.~Legay, and T.~Mens, ``A ground-truth dataset and classification model for detecting bots in github issue and pr comments,'' \emph{Journal of Systems and Software}, vol. 175, p. 110911, 2021.

\bibitem{kruger2019my}
J.~Kr{\"u}ger, M.~Mukelabai, W.~Gu, H.~Shen, R.~Hebig, and T.~Berger, ``Where is my feature and what is it about? a case study on recovering feature facets,'' \emph{Journal of Systems and Software}, vol. 152, pp. 239--253, 2019.

\bibitem{fiechter2021visualizing}
A.~Fiechter, R.~Minelli, C.~Nagy, and M.~Lanza, ``Visualizing github issues,'' in \emph{2021 Working Conference on Software Visualization (VISSOFT)}.\hskip 1em plus 0.5em minus 0.4em\relax IEEE, 2021, pp. 155--159.

\bibitem{zhang2022automatic}
T.~Zhang, I.~C. Irsan, F.~Thung, D.~Han, D.~Lo, and L.~Jiang, ``Automatic pull request title generation,'' \emph{arXiv preprint arXiv:2206.10430}, 2022.

\bibitem{venigalla2024exploratory}
A.~S.~M. Venigalla and S.~Chimalakonda, ``An exploratory study of software artifacts on github from the lens of documentation,'' \emph{Information and Software Technology}, p. 107425, 2024.

\bibitem{venigalla2023docmine}
------, ``Docmine: A software documentation-related dataset of 950 github repositories,'' in \emph{2023 IEEE/ACM 20th International Conference on Mining Software Repositories (MSR)}.\hskip 1em plus 0.5em minus 0.4em\relax IEEE, 2023, pp. 407--411.

\bibitem{du2024evaluating}
X.~Du, M.~Liu, K.~Wang, H.~Wang, J.~Liu, Y.~Chen, J.~Feng, C.~Sha, X.~Peng, and Y.~Lou, ``Evaluating large language models in class-level code generation,'' in \emph{Proceedings of the IEEE/ACM 46th International Conference on Software Engineering}, 2024, pp. 1--13.

\bibitem{liu2024no}
Z.~Liu, Y.~Tang, X.~Luo, Y.~Zhou, and L.~F. Zhang, ``No need to lift a finger anymore? assessing the quality of code generation by chatgpt,'' \emph{IEEE Transactions on Software Engineering}, 2024.

\bibitem{mathews2024test}
N.~S. Mathews and M.~Nagappan, ``Test-driven development and llm-based code generation,'' in \emph{Proceedings of the 39th IEEE/ACM International Conference on Automated Software Engineering}, 2024, pp. 1583--1594.

\bibitem{wu2025empirical}
Y.~Wu, Y.~Wang, Y.~Li, W.~Tao, S.~Yu, H.~Yang, W.~Jiang, and J.~Li, ``An empirical study on commit message generation using llms via in-context learning,'' \emph{arXiv preprint arXiv:2502.18904}, 2025.

\bibitem{ahmed2022few}
T.~Ahmed and P.~Devanbu, ``Few-shot training llms for project-specific code-summarization,'' in \emph{Proceedings of the 37th IEEE/ACM International Conference on Automated Software Engineering}, 2022, pp. 1--5.

\bibitem{ahmed2024automatic}
T.~Ahmed, K.~S. Pai, P.~Devanbu, and E.~Barr, ``Automatic semantic augmentation of language model prompts (for code summarization),'' in \emph{Proceedings of the IEEE/ACM 46th International Conference on Software Engineering}, 2024, pp. 1--13.

\bibitem{wang2024hits}
Z.~Wang, K.~Liu, G.~Li, and Z.~Jin, ``Hits: High-coverage llm-based unit test generation via method slicing,'' in \emph{Proceedings of the 39th IEEE/ACM International Conference on Automated Software Engineering}, 2024, pp. 1258--1268.

\bibitem{ouedraogo2024llms}
W.~C. Ouedraogo, K.~Kabore, H.~Tian, Y.~Song, A.~Koyuncu, J.~Klein, D.~Lo, and T.~F. Bissyande, ``Llms and prompting for unit test generation: A large-scale evaluation,'' in \emph{Proceedings of the 39th IEEE/ACM International Conference on Automated Software Engineering}, 2024, pp. 2464--2465.

\bibitem{hou2024large}
X.~Hou, Y.~Zhao, Y.~Liu, Z.~Yang, K.~Wang, L.~Li, X.~Luo, D.~Lo, J.~Grundy, and H.~Wang, ``Large language models for software engineering: A systematic literature review,'' \emph{ACM Transactions on Software Engineering and Methodology}, vol.~33, no.~8, pp. 1--79, 2024.

\bibitem{chang2024survey}
Y.~Chang, X.~Wang, J.~Wang, Y.~Wu, L.~Yang, K.~Zhu, H.~Chen, X.~Yi, C.~Wang, Y.~Wang \emph{et~al.}, ``A survey on evaluation of large language models,'' \emph{ACM Transactions on Intelligent Systems and Technology}, vol.~15, no.~3, pp. 1--45, 2024.

\bibitem{xia2023automated}
C.~S. Xia, Y.~Wei, and L.~Zhang, ``Automated program repair in the era of large pre-trained language models,'' in \emph{2023 IEEE/ACM 45th International Conference on Software Engineering (ICSE)}.\hskip 1em plus 0.5em minus 0.4em\relax IEEE, 2023, pp. 1482--1494.

\bibitem{macneil2022generating}
S.~MacNeil, A.~Tran, D.~Mogil, S.~Bernstein, E.~Ross, and Z.~Huang, ``Generating diverse code explanations using the gpt-3 large language model,'' in \emph{Proceedings of the 2022 ACM Conference on International Computing Education Research-Volume 2}, 2022, pp. 37--39.

\bibitem{gu2023llm}
Q.~Gu, ``Llm-based code generation method for golang compiler testing,'' in \emph{Proceedings of the 31st ACM Joint European Software Engineering Conference and Symposium on the Foundations of Software Engineering}, 2023, pp. 2201--2203.

\bibitem{dakhel2024effective}
A.~M. Dakhel, A.~Nikanjam, V.~Majdinasab, F.~Khomh, and M.~C. Desmarais, ``Effective test generation using pre-trained large language models and mutation testing,'' \emph{Information and Software Technology}, p. 107468, 2024.

\bibitem{ren2023misuse}
X.~Ren, X.~Ye, D.~Zhao, Z.~Xing, and X.~Yang, ``From misuse to mastery: Enhancing code generation with knowledge-driven ai chaining,'' in \emph{2023 38th IEEE/ACM International Conference on Automated Software Engineering (ASE)}.\hskip 1em plus 0.5em minus 0.4em\relax IEEE, 2023, pp. 976--987.

\bibitem{schafer2023empirical}
M.~Sch{\"a}fer, S.~Nadi, A.~Eghbali, and F.~Tip, ``An empirical evaluation of using large language models for automated unit test generation,'' \emph{IEEE Transactions on Software Engineering}, 2023.

\bibitem{lemieux2023codamosa}
C.~Lemieux, J.~P. Inala, S.~K. Lahiri, and S.~Sen, ``Codamosa: Escaping coverage plateaus in test generation with pre-trained large language models,'' in \emph{2023 IEEE/ACM 45th International Conference on Software Engineering (ICSE)}.\hskip 1em plus 0.5em minus 0.4em\relax IEEE, 2023, pp. 919--931.

\bibitem{su2024distilled}
C.-Y. Su and C.~McMillan, ``Distilled gpt for source code summarization,'' \emph{Automated Software Engineering}, vol.~31, no.~1, p.~22, 2024.

\bibitem{husain2019codesearchnet}
H.~Husain, H.-H. Wu, T.~Gazit, M.~Allamanis, and M.~Brockschmidt, ``Codesearchnet challenge: Evaluating the state of semantic code search,'' \emph{arXiv preprint arXiv:1909.09436}, 2019.

\end{thebibliography}
\bibliographystyle{IEEEtran}

\end{document}